\documentclass{aa}  
\usepackage{graphicx,natbib,psfig}  
\newcommand{\ltsima} {$\; \buildrel < \over \sim \;$}  
\newcommand{\gtsima} {$\; \buildrel > \over \sim \;$}  
\newcommand{\lta} {\lower.5ex\hbox{\ltsima}}  
\newcommand{\gta} {\lower.5ex\hbox{\gtsima}}

\begin{document}  

\title{The host galaxy/AGN connection in nearby early-type galaxies 
\thanks
{Based  on observations obtained at
the  Space  Telescope Science  Institute,  which  is  operated by  the
Association of  Universities for Research  in Astronomy, Incorporated,
under NASA contract NAS 5-26555.}}
\subtitle{Sample selection and hosts brightness profiles}
  
\titlerunning{The host/AGN connection in nearby galaxies }  
\authorrunning{A. Capetti and B. Balmaverde}
  
\author{Alessandro Capetti
\inst{1}
\and    
Barbara Balmaverde \inst{2}}    
   
\offprints{A. Capetti}  
     
\institute{INAF - Osservatorio Astronomico di Torino, Strada
  Osservatorio 20, I-10025 Pino Torinese, Italy\\
\email{capetti@to.astro.it}
\and 
Universit\'a di Torino, Via Giuria 1, I-10125, Torino, Italy\\
\email{balmaverde@ph.unito.it}}

\date{}  
   
\abstract{This is the first of a series of three papers exploring the
connection between  the multiwavelength 
properties of AGNs in nearby early-type galaxies
and the characteristics of their hosts. 
We selected two samples, both with high resolution 5
GHz VLA observations available and providing measurements
down to  1 mJy level, reaching 
radio-luminosities as  low as  10$^{19}$ W  Hz$^{-1}$. 
We focus on the 116 radio-detected galaxies as to boost the fraction
of AGN with respect to a purely optically selected sample.
Here we present the analysis of the optical brightness profiles based on
archival HST images, available for 65 objects. We separate early-type 
galaxies on the  basis of the slope of their  nuclear  brightness
profiles, into core and power-law galaxies following the Nuker's scheme, 
rather than on the traditional morphological classification (i.e. into E and
S0 galaxies). Our sample of AGN candidates is indistinguishable, 
when their brightness profiles
are concerned, from galaxies of similar optical luminosity but
hosting weaker (or no) radio-sources. 
We confirm previous findings that relatively bright radio-sources 
(L$_{\rm{r}} > 10^{21.5}$ W  Hz$^{-1}$) 
are uniquely associated to core galaxies. 
However, below this threshold in radio-luminosity
core and power-law galaxies coexist and 
they do not show any apparent difference in their
radio-properties.  
Not surprisingly, since our sample is 
deliberately biased to favour the inclusion of active galaxies, 
we found a higher fraction of optically nucleated galaxies. 
Addressing the multiwavelength properties of these nuclei will be the 
aim of the two forthcoming papers.
 } \maketitle
  
\section{Introduction}
\label{intro}

It is becoming increasingly clear that most (if not all) galaxies host
a supermassive  black hole (SMBH)  in their centers 
\citep[e.g.][]{kormendy95}.  
The tight relationships  between the SMBH  mass and
the stellar  velocity dispersion \citep{ferrarese00,gebhardt00}
as well as with the mass of  the spheroidal component of
their host galaxies \citep[e.g.][]{marconi03} clearly indicate that
they follow  a common evolutionary  path. As recently  demonstrated by
\citet{heckman04} the  synchronous  growth of  SMBH (traced  by
energy output of AGNs) and galaxies (evidenced by their star formation
rate) is at  work even now, in the nearby  Universe.  But despite this
fundamental  breakthrough  in  our  understanding of  the  SMBH/galaxy
system  and the  fact that  historically the  first evidences  for the
existence of SMBH were provided by  the nuclear activity in form of an
Active Galactic  Nucleus (AGN), we still  lack a clear  picture of the
precise  relationship between  AGN  and host  galaxies.  For  example,
while spiral galaxies preferentially harbour radio-quiet AGN,
early-type   galaxies   host   both   radio-loud   and   radio-quiet
AGN. Similarly, radio-loud AGN  are generally associated with the most
massive SMBH  as there is a  median shift between  the radio-quiet and
radio-loud distribution, but both  distributions are broad and overlap
considerably  \citep[e.g.][]{dunlop03}.

In  the last  decade, thanks  to  HST imaging,  a new  picture of  the
properties of  nearby galaxies also emerged. Nearly  all galaxies have
singular starlight distributions  with surface brightness diverging as
$\Sigma(r)\sim r^{-\gamma},$ with $\gamma>0$ \citep[e.g.][]{lauer95}
and the distribution of cusp slopes in both stellar luminosity density
and  projected surface brightness  is bimodal 
\citep{gebhardt96,faber97}. In some cases  $\gamma$ decreases  only slowly
toward the center  and a steep $\gamma>0.5$ cusp  continues to the HST
resolution  limit;  these  systems  are  classified  as  ``power-law''
galaxies.  In other objects the  projected profile breaks to a shallow
inner  cusp with  $\gamma \leq 0.3$  and  these form  the  class of  ``core
galaxies''.   A small  number of  ``intermediate'' galaxies  have been
also identified  \citep[e.g.][]{ravi01} that have cusp slopes
with $0.3<\gamma<0.5,$  but the ensemble of cusp  slopes of the
spheroidal component of 
galaxies  is still  bimodal. \citeauthor{faber97}  examined how  the
central  structure correlates  with other  galaxy  properties, showing
that luminous  early-type galaxies preferentially  have cores, whereas
most fainter  spheroids have power-law profiles.   Moreover, cores are
slowly  rotating and  have  boxy isophotes,  while  power-laws  rotate
rapidly and are  disky.  This scheme fits nicely  with the revision of
the Hubble sequence proposed by \citet{kormendy96}.

Clearly, these recent developments provide us with a new framework 
in which to explore the connection between host
galaxies and AGNs. In particular, early-type galaxies appear to be the critical
class of objects, in which core and power-law galaxies coexist,
i.e. where the transition between the two profiles classes occur.

This study must be limited to samples of relatively nearby galaxies.
In fact the Nuker classification can only be obtained when the 
nuclear region, potentially associated to a shallow cusp,
can be well resolved. The radius at which the break in
the brightness profiles occurs are in the range 10 pc - 1 kpc; the
most compact cores will be barely resolved already at a distance of 40 Mpc 
(where 10 pc subtend an angle of 0\farcs05)
even in the HST images. Furthermore, high quality radio-images 
are required for an initial selection of AGN candidates. As discussed 
in more details in Sect. \ref{sample} two such large samples of nearby 
early-type galaxies have been already studied at radio-wavelengths
and they will represent the starting point for our analysis.

In order to explore the multiwaveleghts properties of these
candidates AGN we will take advantage on the substantial coverage
supplied by both the HST and Chandra archives when  
nearby galaxies are concerned.  
The high resolution and sensitivity 
of these instruments will enable us to isolate the 
nuclear emission, if present, in both the optical and X-ray band down to
unprecedented low luminosity levels. This will open the possibility
of building three-bands diagnostic diagrams comparing the
radio, optical and X-ray emission, to identify genuine low luminosity
AGN and to explore their spectral energy distributions.
 
The plan of this series of papers is as follows:
in this first part we will define the samples to be examined, collect 
the available HST images, analyze the galaxies brightness profiles
and compare our results with other samples studied in the past. 
In the two forthcoming papers we will firstly discuss the properties of
objects with a core profile, while 
power-law galaxies will be discussed and compared in a third
paper.

We adopt a Hubble constant of H$_{\rm o} = 75$ km s$^{-1}$ Mpc$^{-1}$.

\section{Sample selection and description}
\label{sample}

\begin{table*}
\caption{Galaxies in sample I:
 (1) UGC name, (2) alternative optical identifications, (3) recession velocity 
in km s$^{-1}$, (4) total K band galaxy's magnitude from 2MASS, 
(5) galactic extinction in the V band from NED, (6) 
total 5 GHz radio-flux [mJy] from $^a$ \citet{wrobel91b} or $^b$ \citet{wrobel91a}, (7)
5 GHz radio-core [mJy].}
\label{tabsample1}
\centering
\begin{tabular}{l l r c l  r r}
\hline\hline
Name     & Alt. Name       & V & m$_K$ & Gal. A$_V$ & F$_{\rm radio}$ & F$_{\rm core}$  \\
\hline		      		
UGC~0167 & NGC~0063        & 1197 & 9.257$\pm$0.027 & 0.369 &     3.6$\pm$0.2$^a$ &     \\ 
UGC~0968 & NGC~0524        & 2429 & 7.163$\pm$0.012 & 0.274 &     1.4$\pm$0.1$^a$ &     \\ 
UGC~5292 & NGC~3032        & 1652 & 9.648$\pm$0.024 & 0.056 &     3.7$\pm$0.7$^a$ &     \\ 
UGC~5617 & NGC~3226        & 1410 & 8.570$\pm$0.037 & 0.075 &     3.6$\pm$0.2$^a$ &     \\ 
UGC~5663 & NGC~3245        & 1456 & 7.862$\pm$0.010 & 0.083 &     3.3$\pm$0.2$^a$ &     \\ 
UGC~5902 & NGC~3379,M~105  & 	951 & 6.270$\pm$0.018 & 0.081 &     0.7$\pm$0.1$^a$ &   \\	
UGC~5959 & NGC~3414,ARP~162& 1651 & 7.981$\pm$0.014 & 0.081 &     5.0$\pm$0.2$^a$ &     \\ 
UGC~6153 & NGC~3516        & 2902 & 8.512$\pm$0.027 & 0.140 &    15.5$\pm$1.7$^a$ &     \\ 
UGC~6297 & NGC~3607        & 1022 & 7.276$\pm$0.021 & 0.069 &     2.6$\pm$0.1$^a$ &     \\
UGC~6605 & NGC~3773,MRK~743& 1052 &10.696$\pm$0.063 & 0.089 &     1.6$\pm$0.1$^a$ &     \\ 
UGC~6742 & NGC~3870,MRK~186&  981 &10.820$\pm$0.042 & 0.052 &     4.4$\pm$0.9$^a$ &     \\ 
UGC~6834 & NGC~3928,MRK~190& 1207 & 9.674$\pm$0.025 & 0.065 &     7.6$\pm$1.0$^a$ &     \\ 
UGC~6860 & NGC~3945        & 1500 & 7.526$\pm$0.025 & 0.094 &     1.0$\pm$0.1$^a$ &     \\ 
UGC~6877 & IC~0745,MRK~1308& 1162 &             --- & 0.069 &     0.8$\pm$0.1$^a$ &     \\  
UGC~6946 & NGC~3998        & 1285 & 7.365$\pm$0.010 & 0.053 &      83$\pm$2$^b$   &     \\ 
UGC~6985 & NGC~4026        & 1173 & 7.584$\pm$0.019 & 0.073 &     1.4$\pm$0.1$^a$ &     \\ 
UGC~7005 & NGC~4036        & 1690 & 7.562$\pm$0.019 & 0.078 &     2.9$\pm$0.2$^a$ &     \\ 
UGC~7103 & NGC~4111        & 1006 & 7.553$\pm$0.018 & 0.048 &     2.3$\pm$0.1$^a$ &     \\  
UGC~7142 & NGC~4143        & 1135 & 7.853$\pm$0.009 & 0.042 &     6.7$\pm$0.3$^a$ &     \\ 
UGC~7203 & NGC~4168        & 2392 & 8.437$\pm$0.021 & 0.120 &     4.5$\pm$0.2$^a$ &     \\
UGC~7256 & NGC~4203        & 1263 & 7.406$\pm$0.014 & 0.040 &    12.5$\pm$0.4$^a$ &     \\ 
UGC~7311 & NGC~4233        & 2417 & 8.784$\pm$0.032 & 0.079 &     1.9$\pm$0.1$^a$ &     \\ 
UGC~7329 & NGC~4250        & 2308 & 9.140$\pm$0.027 & 0.069 &     6.4$\pm$1.0$^a$ &     \\ 
UGC~7360 & NGC~4261,3C~270 & 2238 & 7.263$\pm$0.028 & 0.060 &    8300$\pm$400$^b$ & 315 \\ 
UGC~7386 & NGC~4278        &  804 & 7.184$\pm$0.011 & 0.095 &     351$\pm$11$^b$  &     \\
UGC~7494 & NGC~4374,M~84   & 1004 & 6.222$\pm$0.023 & 0.134 &    2800$\pm$100$^b$ & 350 \\
UGC~7515 & NGC~4385,MRK~052& 2172 & 9.776$\pm$0.046 & 0.082 &     3.7$\pm$0.2$^a$ &     \\ 
UGC~7575 & NGC~4435        &  887 & 7.297$\pm$0.016 & 0.100 &     1.2$\pm$0.1$^a$ &     \\ 
UGC~7614 & NGC~4459        & 1294 & 7.152$\pm$0.011 & 0.153 &     0.8$\pm$0.1$^a$ &     \\ 
UGC~7629 & NGC~4472,M~49   &  940 & 5.396$\pm$0.025 & 0.074 &      95$\pm$10 $^b$ &  57 \\
UGC~7654 & NGC~4486,M~87   & 1361 & 5.812$\pm$0.019 & 0.074 &   71900$\pm$400$^b$ &4000 \\
UGC~7718 & NGC~4526        &  665 & 6.473$\pm$0.020 & 0.074 &     3.1$\pm$0.2$^a$ &     \\ 
UGC~7760 & NGC~4552,M~89   &  392 & 6.728$\pm$0.024 & 0.136 &     121$\pm$4$^b$   &     \\
UGC~7797 & NGC~4589        & 2233 & 7.758$\pm$0.023 & 0.093 &    21.0$\pm$0.7$^a$ &     \\
UGC~7878 & NGC~4636        & 1112 & 6.422$\pm$0.035 & 0.092 &    45.0$\pm$9.0$^b$ &   6 \\
UGC~7898 & NGC~4649,M~60   & 1231 & 5.739$\pm$0.021 & 0.088 &    24.0$\pm$2.0$^b$ &  18 \\
UGC~8355 & IC~875          & 3051 &10.296$\pm$0.027 & 0.036 &     0.5$\pm$0.1$^a$ &     \\ 
UGC~8499 & NGC~5198        & 2778 & 8.896$\pm$0.023 & 0.077 &     1.5$\pm$0.1$^a$ &     \\  
UGC~8675 & NGC~5273        & 1316 & 8.665$\pm$0.024 & 0.033 &     1.2$\pm$0.1$^a$ &     \\ 
UGC~8745 & NGC~5322        & 2096 & 7.160$\pm$0.027 & 0.045 &      20$\pm$1$^b$   &  14 \\
UGC~9655 & NGC~5813        & 2025 & 7.413$\pm$0.031 & 0.182 &     2.1$\pm$0.1$^a$ &     \\
UGC~9692 & NGC~5838        & 1450 & 7.581$\pm$0.019 & 0.176 &     2.0$\pm$0.1$^a$ &     \\ 
UGC~9706 & NGC~5846        & 1896 & 6.949$\pm$0.026 & 0.182 &     5.3$\pm$0.3$^a$ &     \\
UGC~9723 & NGC~5866        & 1034 & 6.873$\pm$0.018 & 0.044 &     7.4$\pm$0.3$^a$ &     \\ 
UGC~10656& NGC~6278        & 2981 & 8.994$\pm$0.017 & 0.204 &     1.2$\pm$0.1$^a$ &     \\ 
UGC~12317& NGC~7465,MRK~313& 2046 & 9.542$\pm$0.021 & 0.260 &     3.7$\pm$0.2$^a$ &     \\ 
UGC~12329& NGC~7468,MRK~314& 2164 &11.575$\pm$0.066 & 0.282 &     1.4$\pm$0.1$^a$ &     \\ 
UGC~12759& NGC~7743        & 1725 & 8.418$\pm$0.028 & 0.233 &     2.7$\pm$0.1$^a$ &     \\ 
\hline
\end{tabular}
\end{table*}

\begin{table*}
\caption{Galaxies in sample II:
(1) optical name, (2) alternative optical identifications, (3) recession velocity in km s$^{-1}$,
(4) total K band galaxy's magnitude from 2MASS, (5) galactic extinction in the V band, (6) 
total 5 GHz [mJy] radio-flux from $^a$ \citet{sadler89} 
or $^b$ \citet{slee94}, (7) radio-core flux [mJy].}
\label{tabsample2}
\centering
\begin{tabular}{l l r c l r r }
\hline\hline
Name     & Alt. Name       & V & m$_K$ & Gal. A$_V$ & F$_{\rm radio}$  & F$_{\rm core}$ \\
\hline		      		
NGC~1316 & Fornax A        & 1548 & 5.587$\pm$0.019  & 0.069 & 65800$^b$ &  26 \\
NGC~1380 & E358-G28        & 1621 & 6.869$\pm$0.022  & 0.058 &   1.7$^a$ &     \\
NGC~1399 & E358-G45        & 1217 & 6.306$\pm$0.027  & 0.043 &   342$^b$ &  10 \\
NGC~2328 & E309-G16        &  913 & 9.931$\pm$0.042  & 0.295 &   2.5$^a$ &     \\ 
NGC~3258 & E375-G37        & 2598 & 8.306$\pm$0.024  & 0.279 &    52$^b$ & 3.7 \\
NGC~3268 & E375-G45        & 2600 & 8.153$\pm$0.029  & 0.341 &  22.8$^a$ &     \\
NGC~3557 & E377-G16        & 2857 & 7.203$\pm$0.017  & 0.327 &   299$^b$ &  10 \\
NGC~3706 & E378-G06        & 2827 & 7.902$\pm$0.019  & 0.306 &    19$^b$ & 7.6 \\ 
NGC~4373 & E322-G06        & 3221 & 7.651$\pm$0.022  & 0.265 &  12.7$^a$ &     \\
NGC~4696 & E322-G91        & 2767 & 7.142$\pm$0.024  & 0.376 &  1393$^b$ &  55 \\
NGC~5128 & Cen A           &  390 & 3.942$\pm$0.016  & 0.381 &126000$^b$ &6984 \\
NGC~5419 & E384-G39        & 4044 & 7.518$\pm$0.032  & 0.239 &   390$^b$ &  15 \\
NGC~6958 & E341-G15        & 2566 & 8.400$\pm$0.018  & 0.149 &  18.1$^a$ &     \\
IC~1459  & E406-G30        & 1498 & 6.805$\pm$0.025  & 0.053 &  1016$^b$ &     \\
IC~3370  & E322-G14        & 2773 & 7.858$\pm$0.021  & 0.309 &   6.4$^a$ &     \\ 
IC~4296  & E383-G39        & 3613 & 7.502$\pm$0.021  & 0.204 &  1604$^b$ & 210 \\
IC~4931  & E339-G23        & 5820 & 8.662$\pm$0.030  & 0.234 &   0.9$^a$ &     \\
\hline
\end{tabular}
\end{table*}

For our  purposes we must study  a large sample of  nearby objects for
which  radio observations combining  relatively high  resolution, high
frequency  and sensitivity  are available,  in order  to  minimize the
contribution from radio-emission not related to the galaxy's nucleus
and confusion from background sources.

Two      studies     in      the     literature      fulfill     these
requirements. \citet{wrobel91b}  and \citet{sadler89} presented  5 GHz
VLA radio-images of  two samples of nearby early-type  galaxies with a
resolution  between 3$\arcsec$  and  5$\arcsec$ and  a  flux limit  of
$\sim$  1 mJy.   The two  samples were  selected with  a  very similar
strategy.  More  specifically, \citet{wrobel91a} extracted  a northern
sample  of  galaxies from  the  CfA  redshift survey  \citep{huchra83}
satisfying  the following  criteria: (1)  $\delta_{1950} \geq  0$, (2)
photometric magnitude  B $\leq$ 14; (3) heliocentric  velocity $\leq$ 3000
km s$^{-1}$, and (4) morphological  Hubble type T $\leq$ -1, for a total
number of  216 galaxies.  \citet{sadler89} selected  a southern sample
of E  and S0  galaxies brighter than  magnitude 14, with  $\delta \leq
-32$, (the Parkes sample).   They subsequently selected a subsample
of 116 galaxies  with the further restriction of  $\delta \geq -45$ to
be observable  with the VLA.  The only  substantial difference between
the  two  samples  is  that  \citeauthor{sadler89} did  not  impose  a
distance  limit.   Nonetheless,  the  threshold in  optical  magnitude
effectively limits the  sample to a recession velocity  of $\sim$ 6000
km s$^{-1}$. Consequently their sample has a median recession velocity
larger   by   a   factor   of   $\sim$   2   higher   than   for   the
\citeauthor{wrobel91a}  sample  and  it  contains galaxies  which  are
optically (and radio) brighter by an average factor of 4.

Here we  focus on the  radio-detected sources of both samples,
building a radio-flux limited sample of AGN candidates.  
Having  excluded three
galaxies  from   the  Local  Group   (Leo  I  and  II,   Ursa  Minor),
\citeauthor{wrobel91b} detected 67 galaxies, while 49 were detected by
\citet{sadler89}.   In the HST  public archive  we found  images taken
with WFPC2, NICMOS or  ACS, for a total of 65 objects, 48 and 17 from
the \citeauthor{wrobel91a}  sample (hereafter  Sample I) and  from the
\citeauthor{sadler89}  sample (hereafter  Sample  II) respectively. 
These galaxies form our final sample.

In Tables  \ref{tabsample1} and \ref{tabsample2} we  provide the basic
data for  the selected galaxies  namely the recession velocity
(corrected for infall of the Local Group towards Virgo from the LEDA
database), the K
band  magnitude from  the Two  Micron  All Sky  Survey (2MASS)
(available for  all but one  source), the V band  galactic extinction,
the total radio- luminosity at 5 GHz, the radio-core flux -- 
when this can be isolated in the VLA maps.
   
\section{HST data and surface brightness profiles}
\label{hst}

\begin{figure*}
\centerline{
\psfig{figure=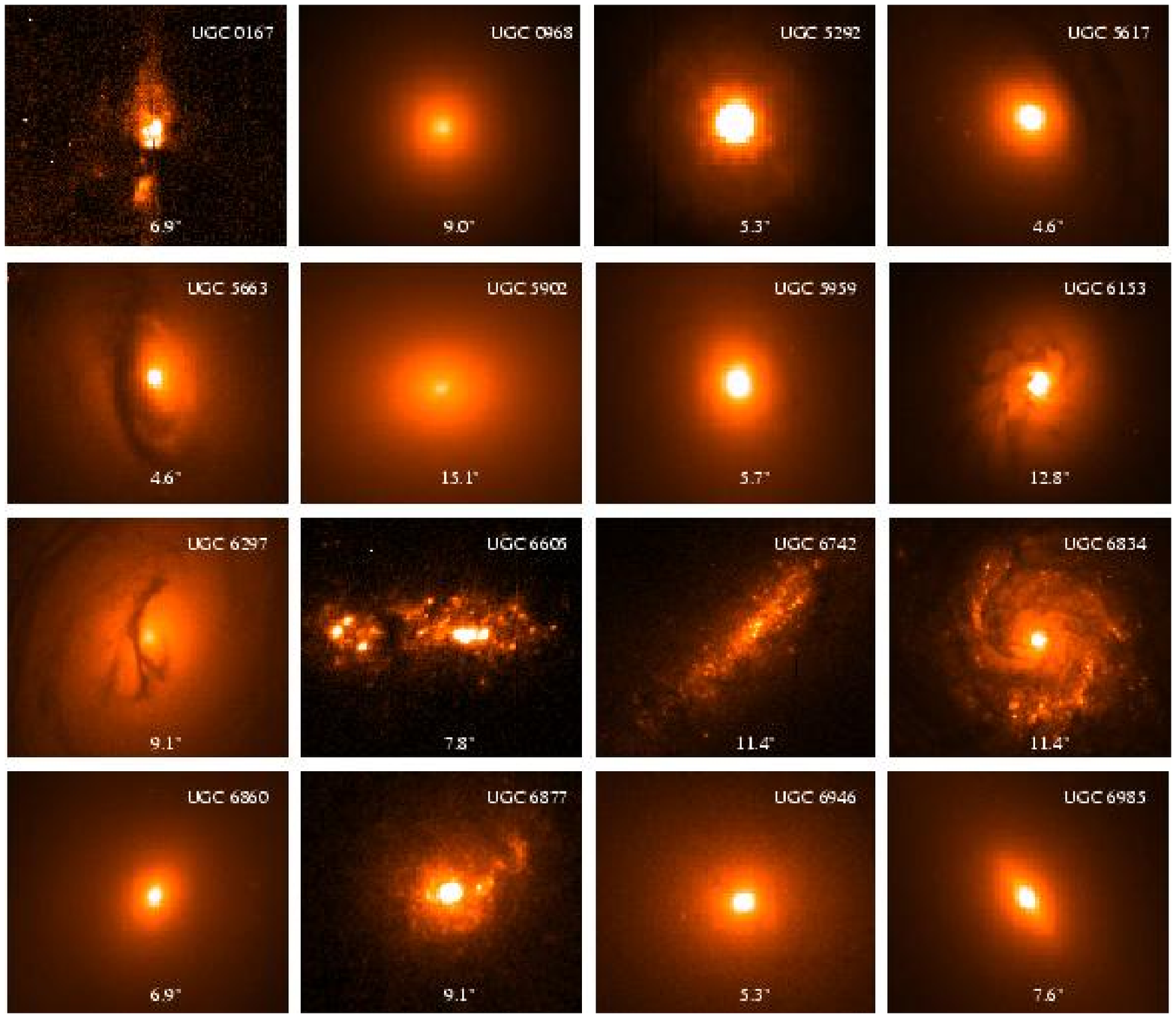,width=1.00\linewidth}
}
\vskip 0.5cm
\caption{\label{w1} HST images of the galaxies of the sample.
The image size is shown in the bottom. The instrument/filter combination is reported 
in Tables \ref{tabobs1} and \ref{tabobs2}} 
\end{figure*}

\begin{figure*}
\centerline{
\psfig{figure=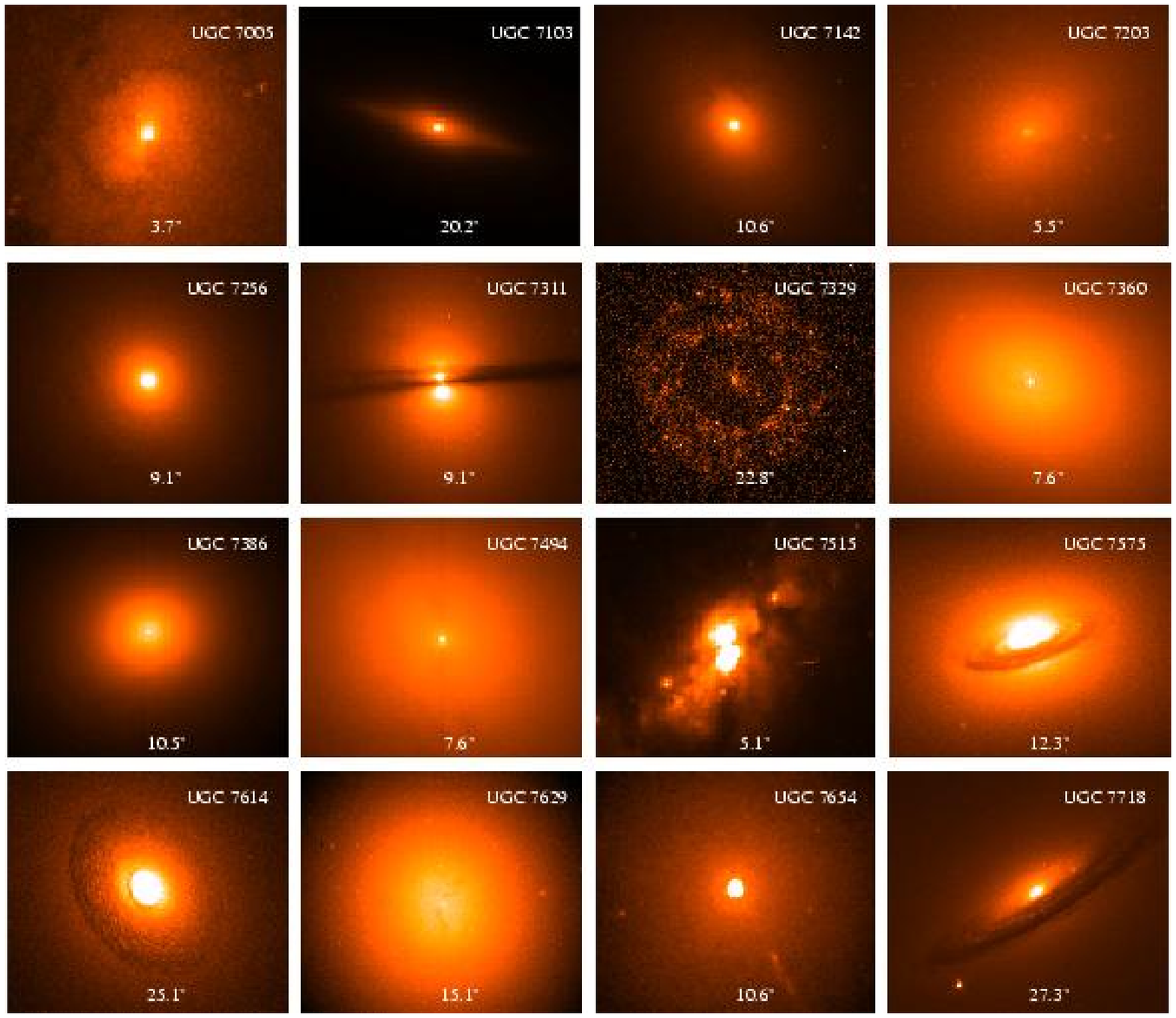,width=1.00\linewidth}
}
\vskip 0.5cm
\caption{\label{w2}  HST images of the galaxies of the sample.}
\end{figure*}

\begin{figure*}
\centerline{
\psfig{figure=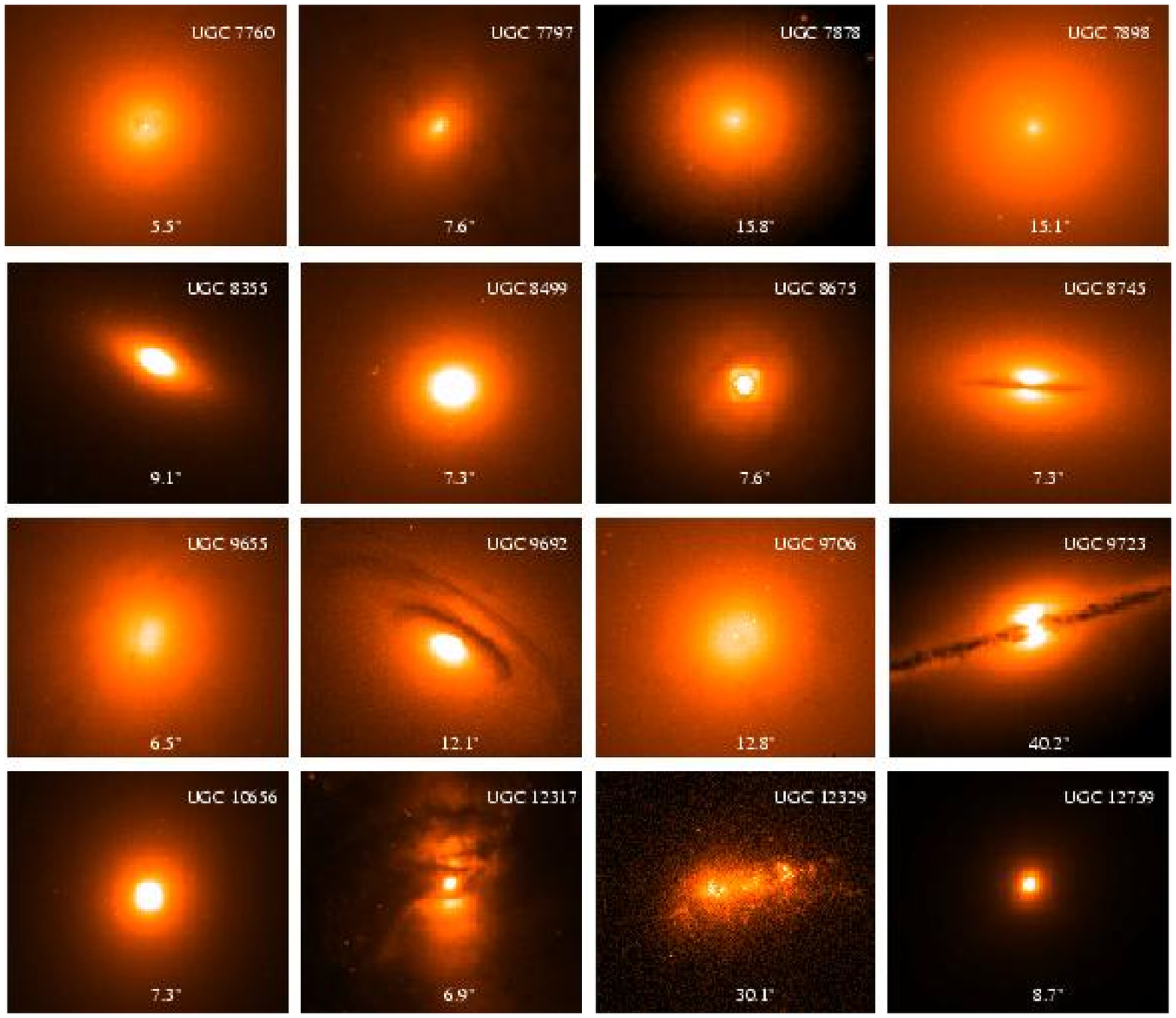,width=1.00\linewidth}
}
\vskip 0.5cm
\caption{\label{w3}  HST images of the galaxies of the sample.}
\end{figure*}

\begin{figure*}
\centerline{
\psfig{figure=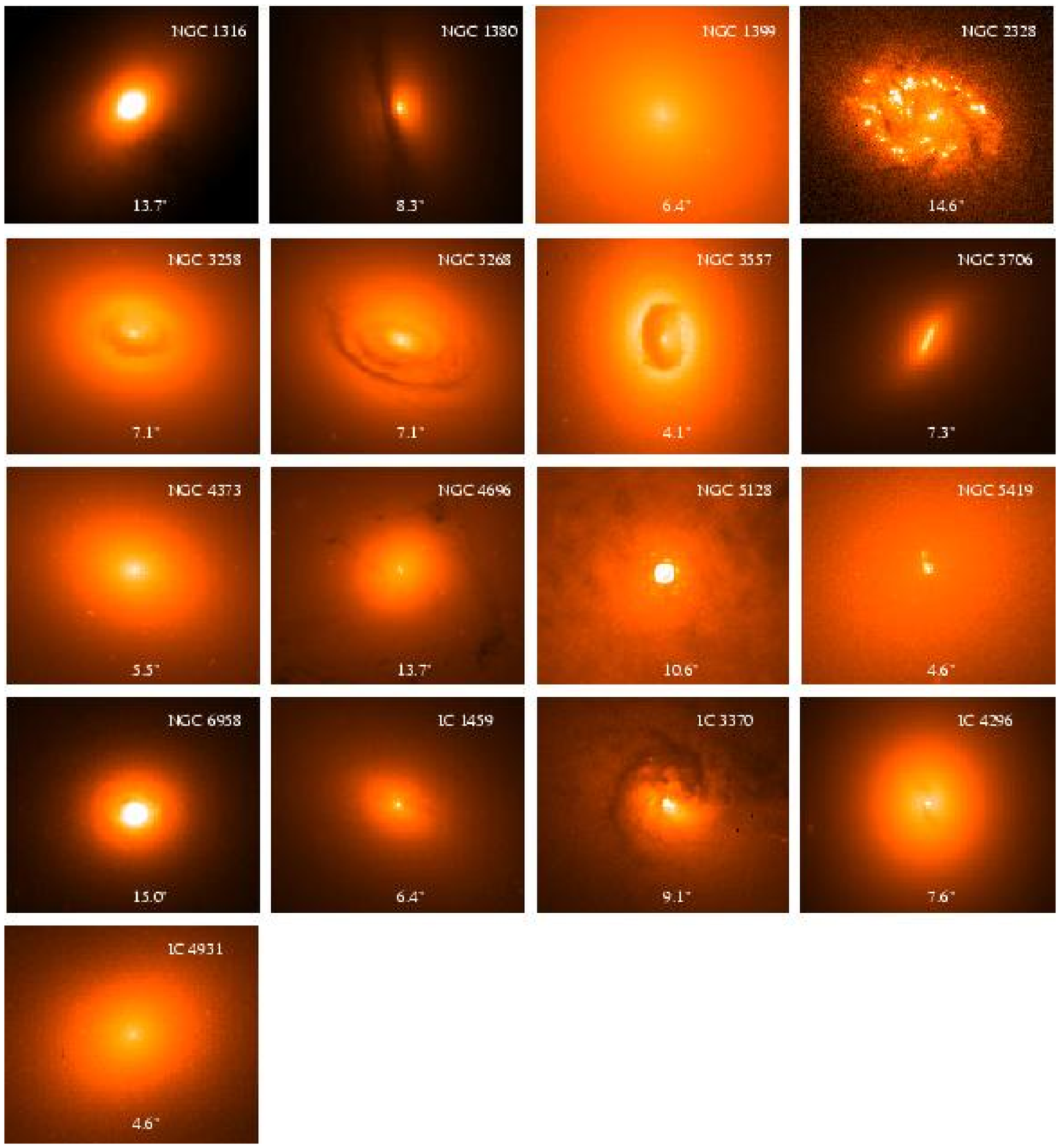,width=1.00\linewidth}
}
\vskip 0.5cm
\caption{\label{s1}  HST images of the galaxies of the sample.}
\end{figure*}

HST  images of  the selected  galaxies were  taken with  a  variety of
instruments and  filters. 
When multiple datasets are  available for a
given object, we  give preference to those obtained  with ACS or WFPC2
in the F814W or F555W filters  to improve the uniformity of the images
used.  However, when significant  dust structures  are present  in the
optical, we used the infrared NICMOS images if present in the archive. 
All data were retrieved from the
public HST archive and were calibrated by
the  standard  On The  Fly  Re-processing  (OTFR)  system.  In  Tables
\ref{tabobs1}   and  \ref{tabobs2}   we  give   the  instrument/filter
combination for the  images used to study the  brightness profile
which are presented
in Fig. \ref{w1} through \ref{s1}.

Several  objects show  peculiar morphologies  on the  HST  scale which
prevent any  attempt to produce  a fit to their  surface brightness.
These include  spiral-like objects,  objects with a  clumpy morphology
suggestive  of star  forming  regions, and galaxies  with severe  dust
obscuration.  In three cases this is probably due to the fact that the
only HST observations available were taken in the ultraviolet with the
F300W  filter.  All  galaxies discarded  at this stage, 
10 + 1 in sample I and II respectively, are 
marked with a dagger in Tables \ref{tabobs1}  and \ref{tabobs2}.
It is interesting to note that all 
11 complex galaxies have $M_K > -22.5$
(with the exception of UGC~7329 with $M_K =-23.0$), substantially
fainter than the median of the regular galaxies ( $M_K =-24.5$) .

When available  we take  the parameters describing  the Nuker  law fit
from the literature.  This  is the case for 21 + 3  objects in the two
sub-samples.  For the remaining 30 objects we used IRAF task 'ellipse'
to  fit elliptical  isophotes to  the  galaxies \citep{jedrzejewski87}.
These include a few objects  for which the classification was based on
pre-refurbishment HST data from \citet{byun96}\footnote{During the
  preparation of this paper \citet{lauer04} published fit to the HST images
of additional 7 objects of our sample. The classification in core and
power-law coincide with those given in the 
Tables \ref{tabobs1} and \ref{tabobs2}  with the only exception of
NGC~3706: while they conclude that this is a core galaxy, we adopt a
more conservative description as a complex profile.}
.

\begin{figure*}
\centerline{
\psfig{figure=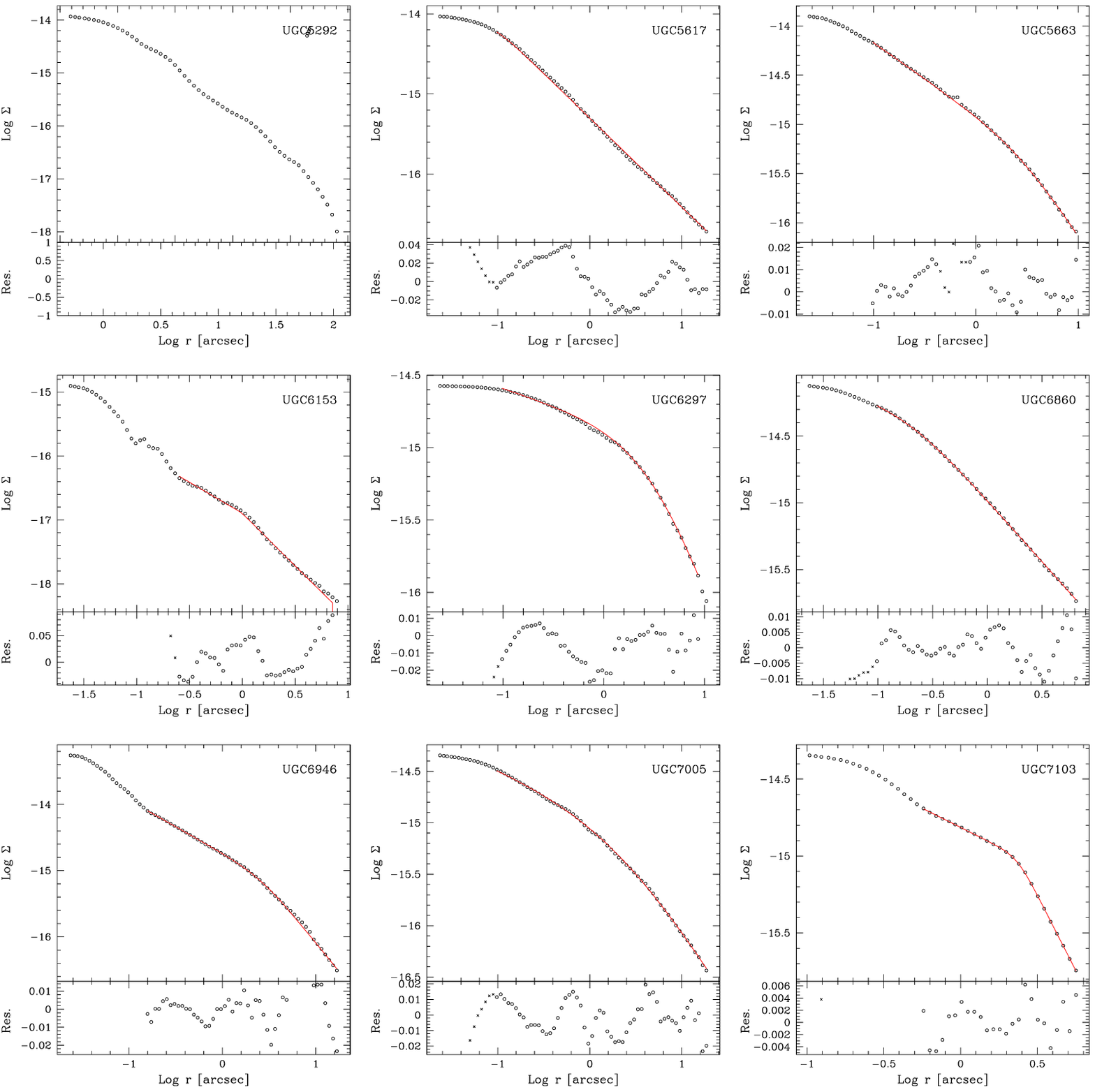,width=1.0\linewidth}}
\caption{\label{profiles1} Surface brightness (in unit of
erg s$^{-1}$ cm$^{-2}$ \AA$^{-1}$) profiles of all galaxies
lacking a core/power-law classification in the literature. Spiral-like
or complex sources are excluded. The solid line represents the best
fit for all objects satisfactorily reproduced with a Nuker law. 
Residuals are presented in the bottom panel. 
Points excluded from the fit have their residuals marked with a cross 
instead of a circle.}
\end{figure*}

\begin{figure*}
\centerline{
\psfig{figure=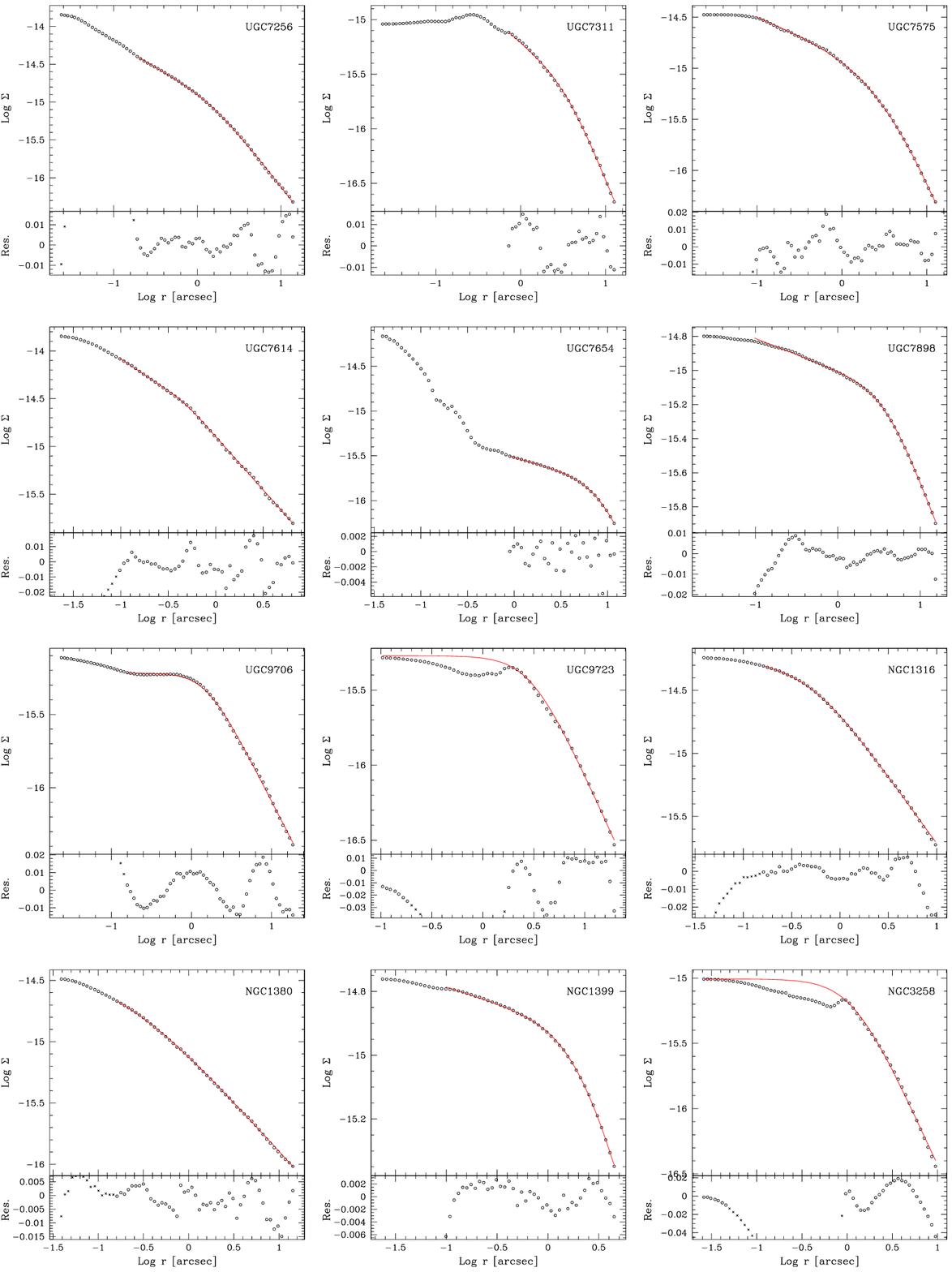,width=1.00\linewidth}}
\caption{\label{profiles2} 
Surface brightness (in unit of
erg s$^{-1}$ cm$^{-2}$ \AA$^{-1}$) profiles of all galaxies
lacking a core/power-law classification in the literature.}
\end{figure*}

\begin{figure*}
\centerline{
\psfig{figure=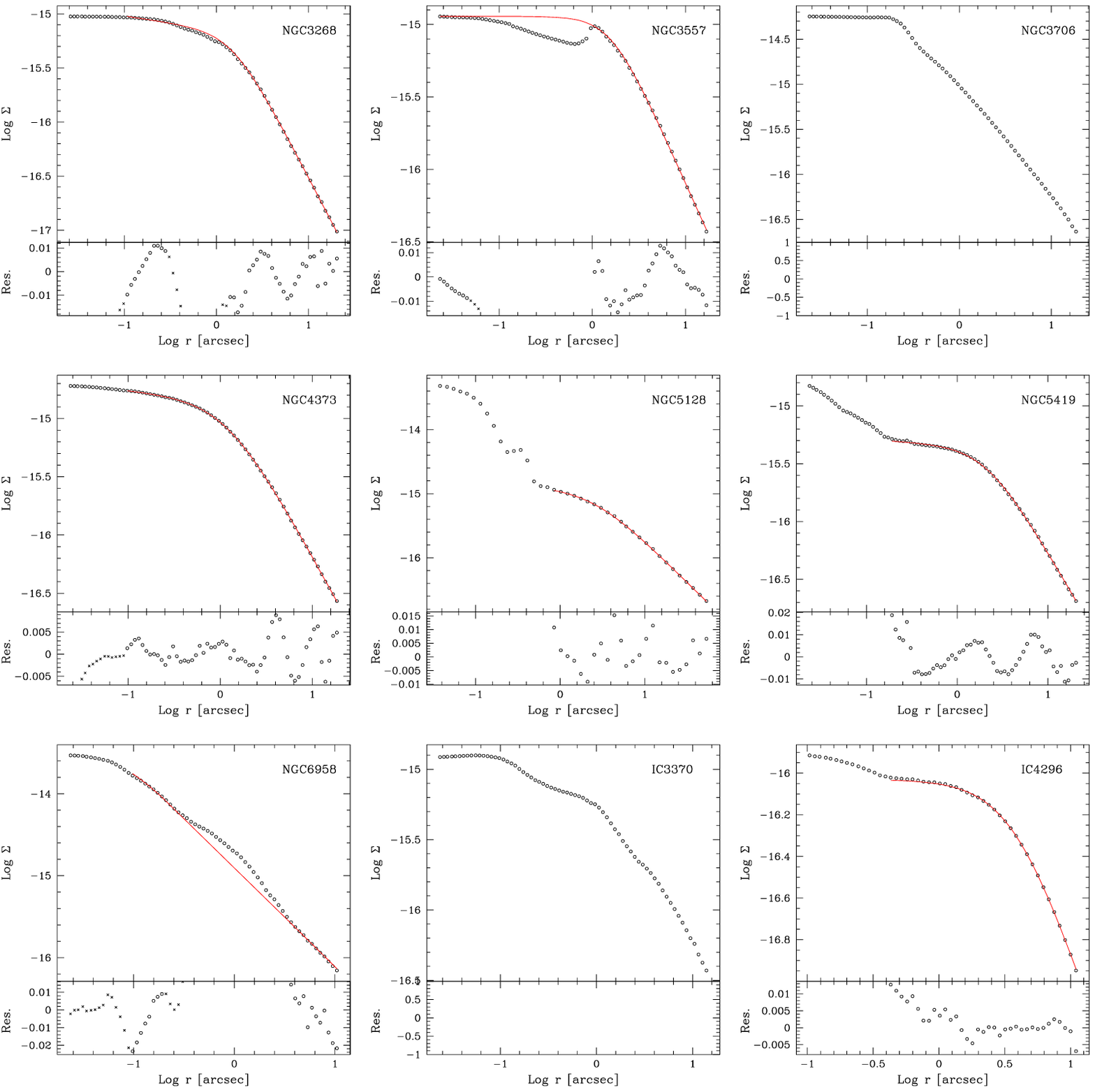,width=1.00\linewidth}}
\caption{\label{profiles3}  
Surface brightness (in unit of
erg s$^{-1}$ cm$^{-2}$ \AA$^{-1}$) profiles of all galaxies
lacking a core/power-law classification in the literature.}
\end{figure*}

Prominent dust features  are often present and they  are masked before
fitting;  in the  radial range  corresponding to  dust  structures, we
usually  preferred to  fix the  isophotes center,  position  angle and
ellipticity to the  values of the last isophote  not affected by dust,
in order  to stabilize the  fit. When the  dust feature cover  a large
azimuthal range (e.g.  in the case of circumnuclear  disks) no masking
can be used  and the resulting profiles will  be necessarily affected.
Similarly  we masked all  background stars  or other  spurious sources
present in the field of view.  The background was measured 
at the edges of images and subtracted. A few objects fill 
the whole available field of view; in these cases
we treated the background level
as an extra free  parameter in the  Nuker fitting. 
The background level is uncorrelated  with all other
parameters  except for the  large scale  slope $\beta$  and it  can be
easily measured and accounted for, particularly in these well
exposed images. The
resulting  one-dimensional major-axis (background subtracted) 
profiles  are presented  in Fig.
\ref{profiles1}  through  \ref{profiles3}.    Errors  in  the  surface
brightness are not be displayed as they rarely exceed the 1\% level.

On these profiles we performed a fit with a Nuker law in the form 

$$ I(r) = I_b 2^{(\beta-\gamma)/\alpha} 
\left({{r_b}\over r}\right)^\gamma 
\left[1 + \left( {r \over {r_b}} \right)^\alpha \right]
^{(\gamma-\beta)/ \alpha} $$

The parameter $\beta$ measures the slope of the outer region of the
brightness profile, $r_b$ is  the break  radius (corresponding to a
brightness $I_b$)  where the profile flattens to a smaller 
slope measured by the
parameter $\gamma$. $\alpha$ sets  the  sharpness  of  the
transition  between the  inner  and outer  profile. 
All models were convolved with the appropriate one dimensional
Point Spread Function profile before they are compared to the data.

We minimized the  value of the $\chi^2$, starting  from a first guess
eye fit and  then exploring the value of $\chi^2$  over a uniform grid
in the  5-dimensional parameter space.   The grid was then  refined in
the region close to the $\chi^2$  minimum to reach an accuracy on each
parameter of 0.01 and finally a downhill simplex minimization algorithm was
applied to convergence. As the behaviour at the smaller
radii might be affected by the details of the Point Spread Function
used for the convolution of the models (such as PSF dependence of
chip location, telescope breathing and so on) we preferred to exclude
points with $r \leq 0\farcs1$ from the fit.
In several objects the HST images show the presence of 
a prominent nuclear point source. 
In these cases we flagged the innermost points of the profile
to an extent which depends on the point source intensity, usually 
the radius where the brightness profile shows the characteristic up-turn.

In  Fig.  \ref{profiles1}  through \ref{profiles3}  we  superposed the
best fit with a  Nuker law to the actual data in  the top panel, while
the  lower panel  presents the  residuals. Regions  masked in the
fitting procedure have their
residuals marked with a cross instead of a circle.

Three galaxies show brightness profiles that cannot be reproduced by a
Nuker law.  These are  marked in Tab.  \ref{tabobs1} and \ref{tabobs2}
as 'complex profiles'. This leaves us with 51 objects. In some cases,
a good fit can only be obtained  by flagging a range of radii; in most
cases this is due to a depression in the brightness profiles that can be
easily associated to dusty regions in the HST images.  In other cases,
localized region  of excess are  found over an otherwise  well defined
Nuker  profile.   Clearly there  is  some  level  of arbitrariness  in
separating complex  profiles from  those only slightly  disturbed.  To
reflect this issue  we mark with an '?' all sources  (6) in which we
flagged a significant portion of the brightness profile.
However,  in  general,  residuals  show large  scale  fluctuations  with
amplitude of  $\leq 1 -  2$ \% over  the whole profile, with  only two
exceptions reaching locally values of 5\%. This implies that overall a
Nuker law reproduces quite well the galaxies profiles.

The parameters derived from the Nuker fits are reported in 
Tab.  \ref{tabobs1} and \ref{tabobs2}. The brightness at the break
radius $\mu_b$ has been converted to the standard Cousin-Johnson system 
and corrected for Galactic extinction. Break radii are in arcseconds.

Sources are separated into power-law and core galaxies adopting the
standard values, i.e. 
$\gamma \leq0.3$ are classified as core-galaxies, while
$\gamma \geq0.5$ are power-law galaxies. We also considered
as power-law galaxies, following
the scheme of \citet{lauer95}, 
all those objects (6) in which no break in the brightness profile is seen 
at the HST resolution limit. More specifically, we included in this
class the galaxies with $r_b \leq 0\farcs2$, as the region
below the break is not sufficiently sampled (when not simply unresolved)
to provide an accurate estimate of $\gamma$.
Potentially, these objects might  
be core galaxies located at a sufficiently large distance so that 
a compact shallow core cannot be resolved. Indeed, there is one core-galaxy
in our sample, UGC~7760 (NGC~4552, M~89) with v=392 km s$^{-1}$, 
with a well resolved core, $r_b = 0\farcs49$ corresponding to 13 pc,
which would have been misclassified with this scheme if it was located
at a larger distance. Thus we prefer
in the remaining of the paper to mark this class differently from
objects with well established power-law profiles.  
In the sample there also 3 galaxies (two among those
analyzed in this work) with 
an intermediate value of $\gamma$. 
See Tab. \ref{nuker} for a summary of the 
breakdown of the objects into the different
classes. 

Since we were forced to use several filters
for the analysis of the surface brightness profile, 
there is the possibility that the description of a given galaxy changes
when observed in different bands. 
However, the comparison of the parameters derived from Nuker
fits on the same galaxy imaged with HST in different filters 
\citep{ravi01} indicates that,
although differences are present, the classification into power-law
and core galaxies is independent on wavelength.
Nonetheless, we prefer to explore 
this issue in more detail for our sample.
The images 
used in this work range from the V to the H band. We here consider
the 21 galaxies for which images taken in both these bands are
available, to test the robustness of the
Nuker classification over the largest breadth of wavelength. 
Nine objects must be discarded since their V band images are
saturated (3), heavily affected by dust (4) or as the presence of a
single image in the archive prevents cosmic-ray rejection (2). 
Of the 12 remaining objects, brightness profiles of 
11 galaxies have been already presented in the literature 
(see the articles cited in Table 3, \citet{lauer04} and 
\citet{faber97}), leaving to us only the analysis of the V band image
of UGC~7494 (M~84).
No discordant Nuker classification has been found, further supporting
its independence on the observing band.

\begin{table*}
\caption{Nuker parameters for sample I}
\label{tabobs1}
\centering
\begin{tabular}{l l c c c c c c l}
\hline\hline
Name      &Image&$\alpha$&$\beta$&$\gamma$&$r_b$&$\mu_b$&   Ref. & Classification\\
\hline	      	 	         
UGC~0167  & WFPC2/F300W  & \multicolumn{5}{l}{$\dagger$  Complex}                  &  (1)   &     ---    \\
UGC~0968  & NICMOS/F160W &     0.68  &     1.69   &    0.03   &    1.41  &    13.73&  (2)   & Core       \\
UGC~5292  & NICMOS/F160W & \multicolumn{5}{l}{Complex profile}                     &  (1)   &     ---    \\
UGC~5617  & WFPC2/F702W  &     ---   &     1.11   &    ---    & $<$0.20  & $<$15.17&  (1)   & Power-law* \\
UGC~5663  & WFPC2/F702W  &     1.87  &     1.59   &    0.73   &    2.69  &    16.48&  (1)   & Power-law  \\
UGC~5902  & NICMOS/F160W &     1.82  &     1.45   &    0.18   &    1.58  &    12.80&  (2)   & Core       \\
UGC~5959  & WFPC2/F702W  &     1.40  &     1.45   &    0.83   &    1.72  &    16.46&  (3)   & Power-law  \\
UGC~6153  & NICMOS/F160W &    42.60  &     1.64   &    0.97   &    1.61  &    13.97&  (1)   & Power-law? \\
UGC~6297  & WFPC2/F814W  &     1.51  &     1.97   &    0.23   &    3.22  &    15.86&  (1)   & Core       \\
UGC~6605  & WFPC2/F300W  & \multicolumn{5}{l}{$\dagger$   Complex}                 &  (1)   &     ---    \\ 
UGC~6742  & WFPC2/F814W  & \multicolumn{5}{l}{$\dagger$   Complex}                 &  (1)   &     ---    \\ 
UGC~6834  & WFPC2/F606W  & \multicolumn{5}{l}{$\dagger$   Spiral}                  &  (1)   &     ---    \\ 
UGC~6860  & WFPC2/F814W  &     ---   &     0.92   &    ---    & $<$0.20  & $<$14.57&  (1)   & Power-law* \\
UGC~6877  & NICMOS/F160W & \multicolumn{5}{l}{$\dagger$   Complex}                 &  (1)   &     ---    \\
UGC~6946  & WFPC2/F547M  &     2.94  &     1.75   &    0.80   &    2.61  &    16.76&  (1)   & Power-law  \\
UGC~6985  & NICMOS/F160W &     0.88  &     1.50   &    0.68   &    0.84  &    12.40&  (2)   & Power-law  \\
UGC~7005  & WFPC2/F547M  &     0.89  &     1.46   &    0.36   &    1.98  &    17.08&  (1)   & Interm.    \\
UGC~7103  & NICMOS/F160W &    10.50  &     1.97   &    0.50   &    2.35  &    12.90&  (1)   & Power-law  \\
UGC~7142  & NICMOS/F160W &     1.26  &     2.18   &    0.59   &    3.11  &    14.26&  (2)   & Power-law  \\
UGC~7203  & WFPC2/F702W  &     1.43  &     1.39   &    0.17   &    2.02  &    17.45&  (3)   & Core       \\
UGC~7256  & WFPC2/F814W  &     2.13  &     1.41   &    0.62   &    1.50  &    15.33&  (1)   & Power-law  \\
UGC~7311  & WFPC2/F702W  &     2.49  &     1.95   &    0.64   &    3.43  &    18.11&  (1)   & Power-law? \\
UGC~7329  & WFPC2/F300W  & \multicolumn{5}{l}{$\dagger$   Complex}                 &  (1)   &     ---    \\
UGC~7360  & NICMOS/F160W &     2.38  &     1.43   &    0.00   &    1.62  &    13.58&  (2)   & Core       \\
UGC~7386  & NICMOS/F160W &     1.63  &     1.39   &    0.02   &    0.97  &    12.80&  (2)   & Core       \\
UGC~7494  & NICMOS/F160W &     2.15  &     1.50   &    0.13   &    2.39  &    13.35&  (2)   & Core       \\
UGC~7515  & WFPC2/F606W  & \multicolumn{5}{l}{$\dagger$   Complex}                 &  (1)   &     ---    \\
UGC~7575  & WFPC2/F814W  &     1.25  &     2.13   &    0.33   &    3.45  &    16.22&  (1)   & Interm.    \\
UGC~7614  & WFPC2/F814W  &     6.51  &     1.12   &    0.66   &    0.47  &    13.97&  (1)   & Power-law  \\
UGC~7629  & NICMOS/F160W &     1.89  &     1.29   &    0.04   &    2.63  &    13.52&  (2)   & Core       \\
UGC~7654  & NICMOS/F160W &     2.56  &     2.78   &    0.28   &    9.41  &    15.51&  (1)   & Core       \\
UGC~7718  & WFPC2/F814W  & \multicolumn{5}{l}{$\dagger$   Dusty}                   &  (1)   &     ---    \\
UGC~7760  & WFPC2/F555W  &     3.03  &     1.06   &    0.00   &    0.49  &    15.24&  (4)   & Core       \\
UGC~7797  & NICMOS/F160W &     1.09  &     1.18   &    0.11   &    0.21  &    12.05&  (2)   & Core       \\
UGC~7878  & NICMOS/F160W &     1.69  &     1.56   &    0.13   &    3.44  &    14.60&  (2)   & Core       \\
UGC~7898  & WFPC2/F555W  &     2.44  &     1.27   &    0.19   &    3.83  &    16.95&  (1)   & Core       \\
UGC~8355  & WFPC2/F702W  &     0.85  &     1.81   &    1.07   &    1.87  &    17.65&  (3)   & Power-law  \\
UGC~8499  & WFPC2/F702W  &     2.61  &     1.13   &    0.23   &    0.16  &    14.79&  (3)   & Power-law* \\
UGC~8675  & NICMOS/F160W &     7.03  &     1.32   &    0.37   &    0.65  &    13.96&  (2)   & Interm.    \\
UGC~8745  & WFPC2/F814W  &     1.11  &     1.45   &    0.00   &    0.68  &    14.59&  (4)   & Core       \\
UGC~9655  & WFPC2/F702W  &     1.90  &     1.33   &    0.00   &    0.73  &    15.77&  (3)   & Core       \\
UGC~9692  & NICMOS/F160W &     4.35  &     2.57   &    0.93   &    4.35  &    14.84&  (2)   & Power-law  \\
UGC~9706  & WFPC2/F702W  &     2.62  &     1.07   &    0.00   &    1.52  &    17.12&  (1)   & Core       \\
UGC~9723  & NICMOS/F160W &     2.50  &     1.51   &    0.00   &    2.99  &    13.61&  (1)   & Core?      \\
UGC~10656 & WFPC2/F702W  &     0.76  &     1.62   &    0.55   &    0.60  &    15.69&  (3)   & Power-law  \\
UGC~12317 & WFPC2/F606W  & \multicolumn{5}{l}{$\dagger$   Complex}                 & (1)    &     ---    \\
UGC~12329 & WFPC2/F814W  & \multicolumn{5}{l}{$\dagger$   Complex}                 & (1)    &     ---    \\
UGC~12759 & NICMOS/F160W &     5.36  &     1.38   &    0.50   &    0.16  &    11.59&  (2)   & Power-law* \\
\hline	      	   
\hline
\end{tabular}

References: (1) this work, (2) Ravindranath et al. 2001, 
(3) Rest et al. 2001, (4) Carollo et al. 1997

* power-law galaxies with $r_b \leq 0.2$; 
? tentative classification. 
\end{table*}

\begin{table*}
\caption{Nuker parameters for sample II}
\label{tabobs2}
\centering
\begin{tabular}{l l c c c c c c l}
\hline\hline
Name      &Image&$\alpha$&$\beta$&$\gamma$&$r_b$&$\mu_b$&   Ref. & Classification\\
\hline	      	 	         
NGC~1316 & NICMOS/F160W&     1.88  &     1.06   &    0.13   &    0.52  &    11.54 & (1) & Core       \\
NGC~1380 & NICMOS/F160W&     ---   &     0.81   &    ---    &$<$0.20& $<$12.08  & (1)   & Power-law* \\
NGC~1399 & WFPC2/F814W &     2.04  &     1.21   &    0.10   &    2.19  &    15.37 & (1) & Core       \\
NGC~2328 & WFPC2/F814W & \multicolumn{5}{l}{$\dagger$   Spiral}                   & (1) & ---        \\
NGC~3258 & ACS/F814W   &     2.10  &     1.51   &    0.00   &    1.15  &    15.59 & (1) & Core?      \\
NGC~3268 & ACS/F814W   &     2.49  &     1.64   &    0.13   &    1.51  &    15.87 & (1) & Core       \\
NGC~3557 & WFPC2/F555W &     2.62  &     1.44   &    0.00   &    1.60  &    16.38 & (1) & Core?      \\
NGC~3706 & WFPC2/F555W & \multicolumn{5}{l}{Complex profile}                      & (1) & ---        \\
NGC~4373 & WFPC2/F814W &     1.69  &     1.45   &    0.08   &    1.19  &    15.25 & (1) & Core       \\
NGC~4696 & WFPC2/F814W &     6.63  &     0.86   &    0.10   &    1.40  &    16.36 & (5) & Core       \\
NGC~5128 & NICMOS/F222M&     1.68  &     1.30   &    0.10   &    2.56  &    11.98 & (1) & Core       \\
NGC~5419 & WFPC2/F555W &     2.01  &     1.35   &    0.05   &    2.11  &    17.56 & (1) & Core       \\
NGC~6958 & WFPC2/F814W &     ---   &     1.22   &    ---    & $<$0.20  & $<$13.63 & (1) & Power-law*?\\
IC~1459  & WFPC2/F814W &     0.81  &     2.04   &    0.02   &    1.82  &    15.15 & (4) & Core       \\
IC~3370  & WFPC2/F555W & \multicolumn{5}{l}{Complex profile}                      & (1) & ---        \\
IC~4296  & NICMOS/F160W&     2.14  &     1.90   &    0.00   &    1.44  &    13.89 & (1) & Core       \\
IC~4931  & WFPC2/F814W &     2.08  &     1.46   &    0.09   &    0.99  &    15.74 & (5) & Core       \\
\hline	      	   
\hline
\end{tabular}

References: (1) this work, (4) Carollo et
al. 1997, (5) Laine et al. 2003

* power-law galaxies with $r_b \leq 0.2$; 
? tentative classification. 
\end{table*}

\begin{table}
\caption{Nuker classification summary}
\label{nuker}
\centering
\begin{tabular}{l c c}
\hline\hline
Class         & \# Sample I & \# Sample II \\
\hline	      	 	         
Core galaxies        &  16 + 1? & 10 + 2? \\
Power-law galaxies   &  10 + 5* + 2? &  1* + 1? \\
Interm. galaxies     &   3      &  0 \\
Complex profiles     &   1      &  2 \\
Complex morphologies &  10      &  1 \\
\hline
Total                &  48 & 17 \\	      	   
\hline
\end{tabular}
\\ 
? tentative classification;
* power-law galaxies with $r_b \leq 0.2$ 
\end{table}

\begin{figure}
\centerline{
\psfig{figure=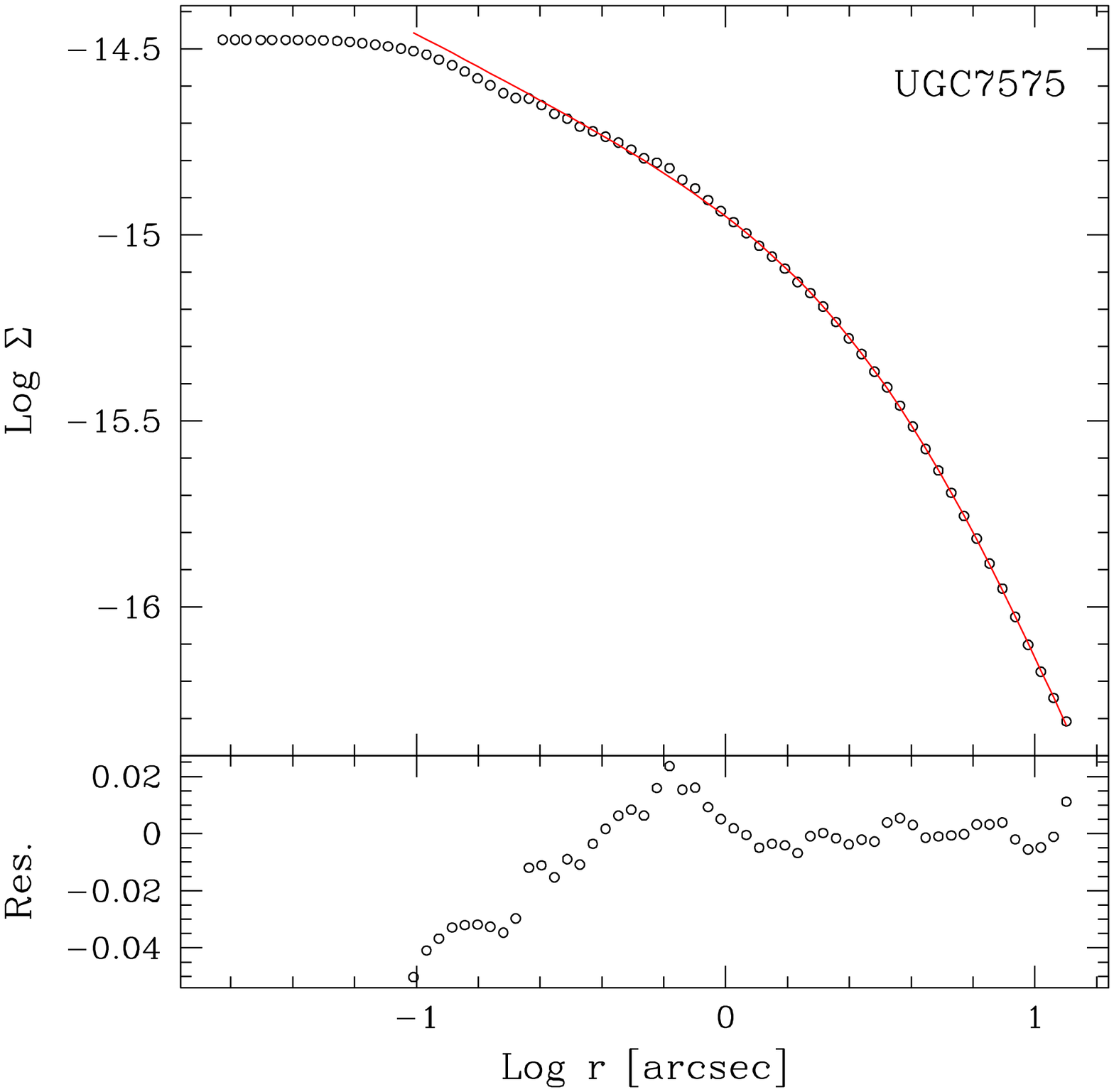,width=1.00\linewidth}
}
\vskip 0.5cm
\caption{\label{ugc7575} 
Nuker fit to  UGC~7575 forcing the value of $\gamma$  to be 0.1 larger
than the  best fit  value of 0.33,  while leaving al  other parameters
free to vary.  This representative example is used  to illustrate that
the error on this parameter is in general smaller than 0.1.
}
\end{figure}

As discussed  by several authors, a  formal estimate of  errors on the
resulting Nuker parameters can not be performed \citep[e.g.][]{byun96}.
Since  we are mostly  interested in  separating galaxies  with shallow
cusps from those  with steep cusps, we explored  how the variations of
$\gamma$  are  reflected  in  the   fit  and  in  the  residuals. 
We focus on UGC~7575,
one of the  two galaxies for which we found  an intermediate value of
$\gamma = 0.33$. Its  best fit (see Fig.\ref{profiles2}) 
shows  residuals  always  smaller than  $\pm
1.5\%$, quite typical for all  fitted galaxies.
In Fig. \ref{ugc7575} we  present the Nuker fit obtained forcing 
$\gamma$ to 0.43, a value 0.1 larger with respect  to the best  fit,
leaving all other parameters free to vary. Large and systematic
residuals  appear toward the center, 
reaching 5\%  at 0\farcs1.  
Although this  does not
provide us with a precise  error estimate,  it  indicates that  the error  on
$\gamma$  are generally   significantly  smaller  than  0.1,  in
agreement  with the  analysis performed  by  \citeauthor{byun96} This
implies that  except for a few border-line  objects our classification
is robust.

\section{Discussion}
\label{discussion}

\begin{figure}
\psfig{figure=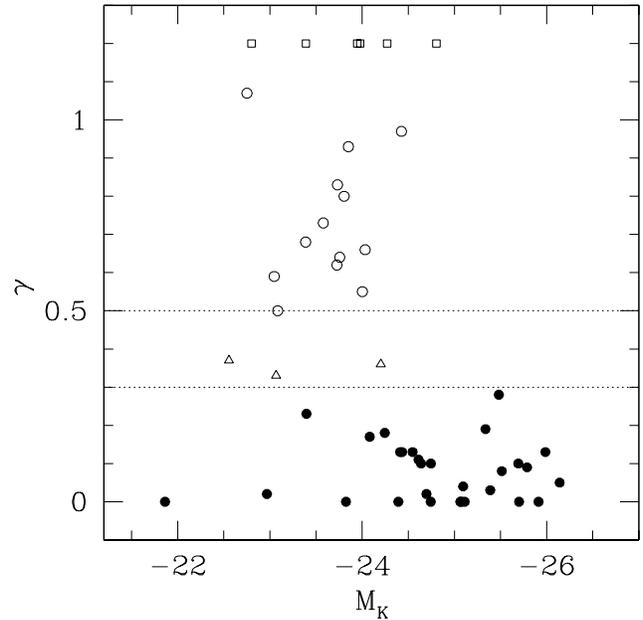,width=1.00\linewidth}
\caption{\label{mkgamm} Logarithmic inner slope derived
from the Nuker fit vs galaxy's K band magnitude. Galaxies belonging to
sample I and II are separated on the basis of the value of
$\gamma$. Core galaxies have $\gamma \leq 0.3$ (filled circles), 
intermediate galaxies
have  $0.3<\gamma<0.5$ (empty triangles), 
power-law galaxies have $\gamma \geq 
0.5$ (empty circles). Power-law galaxies in
which no break in the brightness profile is seen above the
resolution
limit are located arbitrarily at
  $\gamma=1.2$ and are marked with empty squares.  }
\end{figure}

\begin{figure}
\psfig{figure=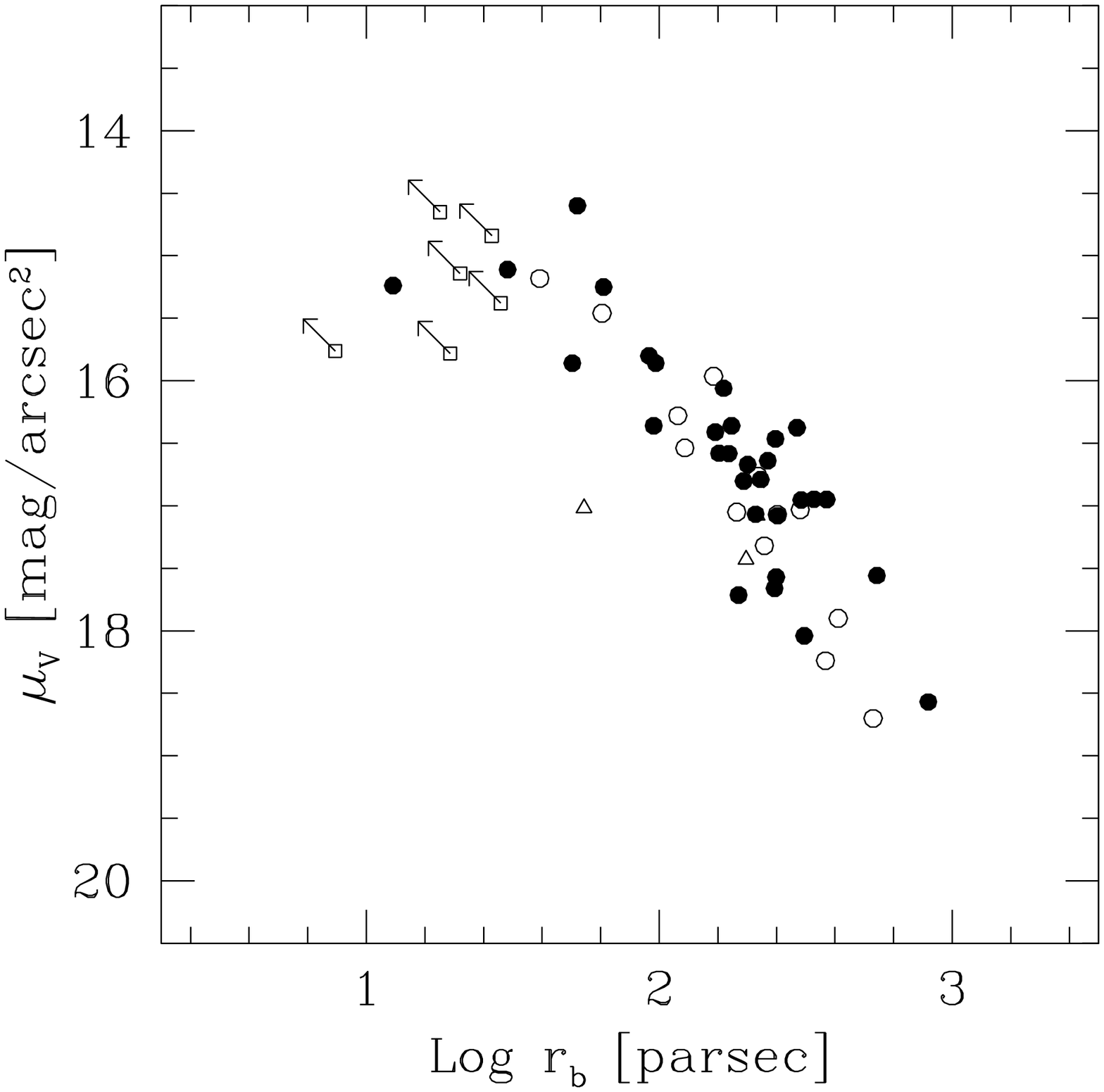,width=1.00\linewidth}
\caption{\label{mubrb} Surface brightness at the break radius $\mu_b$
vs break radius  $r_b$. Symbols as in Fig. \ref{mkgamm},
i.e. filled circles for core galaxies, empty circles for power-laws,
triangles for intermediate galaxies, squares for 
galaxies with no break in the brightness profile.}
\end{figure}

We are now in the position to explore how the multiwavelength properties 
of the galaxies of our sample are related to the brightness profiles 
and how they compare with
results obtained on purely optically selected samples.
In Fig. \ref{mkgamm} we plot the absolute K band magnitude M$_K$ derived 
from 2MASS measurements\footnote{We
preferred to use the K band luminosities, despite all previous studies
have used optical measurements, since the near-infrared light is a
better tracer of the stellar mass and less subject to the effects of
extinction. Furthermore, 2MASS provides us with a uniform and accurate
set of measurements. The conversion to e.g. the V band can be obtained
using the average color for E galaxies, V-K = 3.3, see
\citet{mannucci01}.} with the asymptotic slope $\gamma$.
We recover the already known difference between core and
power-law galaxies in terms of absolute magnitude. The most luminous galaxies 
are exclusively associated to core profiles. Nonetheless, core-galaxies are observed
over a broad range of luminosity 
extending from $M_K \sim -26.5$ down to $M_K \sim -21.8$, more
  than two magnitudes above and
below the break in the luminosity function, M$^*_K$=-24.3 \citep{huang03}. 
Leaving aside the objects in
which no break in the brightness profile is observed, 
the brightest power-law galaxy has $M_K = -24.2$. Therefore, there is
a substantial overlap between the two classes as far as their
luminosity is concerned. 

The well defined correlation 
between break radius and surface brightness at the break, 
which have been described as a 
fundamental plane for the galaxies cores, is also recovered 
(see Fig. \ref{mubrb})\footnote{We converted surface brightness in the
different filters into the V band using the average colors for E/S0
tabulated by \citet{mannucci01}}. The comparison with previous works
(see e.g. Fig. 8 in \citeauthor{faber97} and 
the compilation provided by De Ruiter et al. 2005) 
indicate that the  $r_b$ vs $\mu_b$ relationship 
we obtain is in very good agreement
from that seen in the other samples studied, purely optically selected.
We conclude that our sample
does not reflect any significant difference in the host galaxies
with respect to the overall population of early-type galaxies, 
despite it has been selected imposing a minimum radio-flux level.

The only significant difference
is the higher fraction of nucleated galaxies;
this fraction varies considerably in previous studies e.g. from  
$\sim$ 15\% in \citeauthor{laine03} to $\sim$ 40\%
in \citeauthor{lauer04}
Although we defer  
the analysis of the nuclear properties to the forthcoming papers, 
we note that in our sample we detected an optical (or infrared)
nucleus in $\sim 65\%$ of the objects. 
This is not at all unexpected, since our sample is 
deliberately biased to favour the inclusion of active galaxies.

\begin{figure}
\psfig{figure=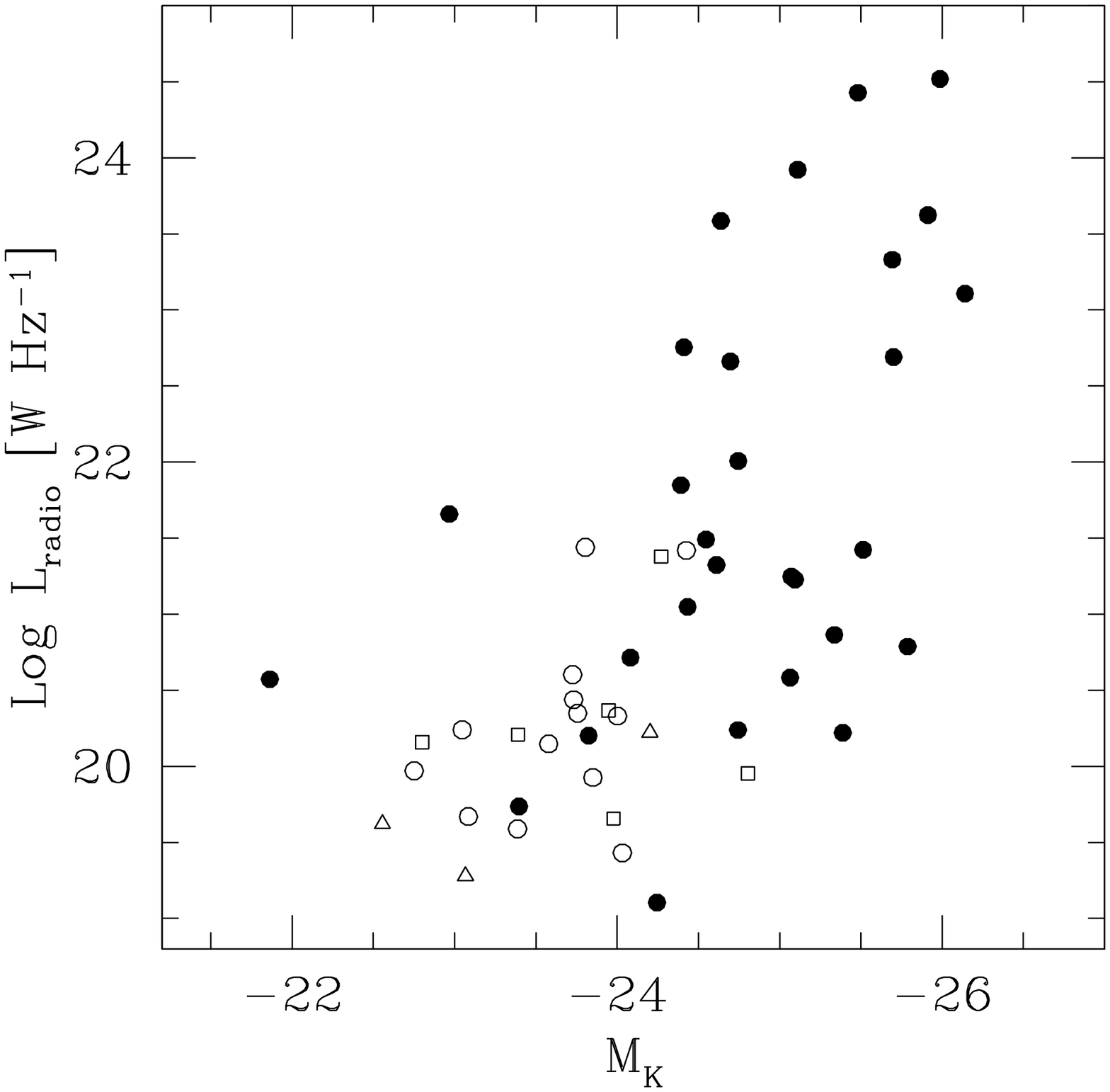,width=1.00\linewidth}
\caption{\label{mkr} Host galaxies magnitude vs radio-luminosity.
Symbols: filled circles for core galaxies, empty circles for power-laws,
triangles for intermediate galaxies, squares for 
galaxies with no break in the brightness profile.}
\end{figure}

\begin{figure}
\psfig{figure=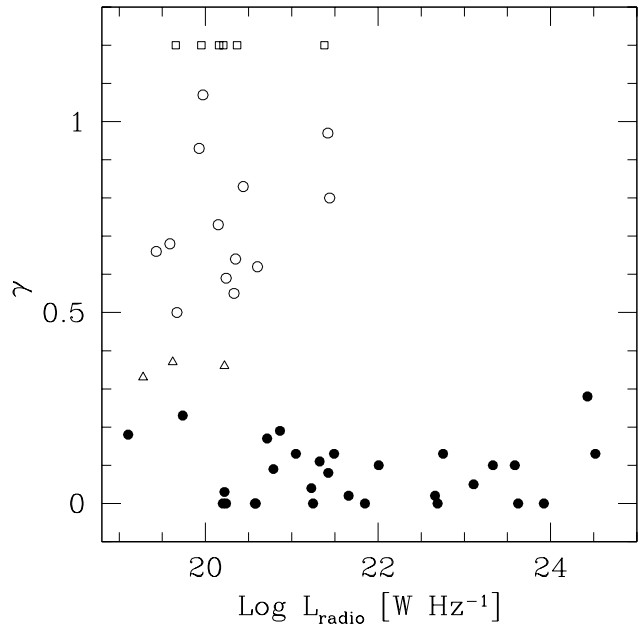,width=1.00\linewidth}
\caption{\label{gammr}  Logarithmic inner slope vs radio-luminosity.
Symbols: filled circles for core galaxies, empty circles for power-laws,
triangles for intermediate galaxies, squares for 
galaxies with no break in the brightness profile.}
\end{figure}

Concerning the relationship
between host's magnitude and radio-power this is visually presented in
Fig. \ref{mkr}. Several authors in the past discussed this
relationship \citep[e.g.][]{auriemma77} 
that can be described as a trend for which brighter galaxies 
have a higher probability to be strong
radio-emitters with respect to less luminous galaxies. 
By separating core and power-law galaxies a clearer picture emerges. 
Only core-galaxies are radio-emitters at level larger than
$L_r > 2.5 \times 10^{21}$ W Hz$^{-1}$. This is in line with
the result found by De Ruiter et al. (2005); their analysis of the 
brightness profiles of HST images of radio-galaxies extracted from the
B2 sample, as well as from nearby 3C sources, shows that all host
galaxies of these relatively powerful radio-sources have a core profile.
Fig. \ref{gammr} provides us with a different view of
this same effect, by comparing $\gamma$ with $L_r$. 
It also shows that, when the two classes
are separated, there is no dependence
of the radio-luminosity on the value of $\gamma$, e.g. 
smaller values of $\gamma$ in core galaxies are not associated to 
the brightest radio-sources. 

However, below this threshold,
the two populations of early-type galaxies cannot be readily 
differentiated. 
We are left with the ambiguity on what is the
driving mechanism at the origin of this threshold in radio-luminosity
for power-law galaxies: it can be due to an intrinsic difference
with respect to core galaxies, but it might also be related to
the difference in the host galaxy's luminosity, since no strong
radio-source ($L_r > 10^{22}$ W Hz s$^{-1}$) is associated to a host
with $M_K > -24$, regardless of the profile class.

Considering separately the two classes,
the (optically) brightest core galaxies are indeed 
associated to the strongest radio-sources. However, this effect cannot
be simply described as a correlation between  L$_r$ and  M$_K$ 
since they show a very large spread,
covering as much as 3 dex in radio-luminosity at a given
magnitude. Furthermore, this trend does not extend to the lowest
host magnitudes. 

If we now consider the galaxies with intermediate nuclear slope 
and those in which no break in the 
brightness profile is seen (marked as triangles and squares
respectively in Fig. \ref{mkr}) they are all relatively faint 
radio-sources, below the threshold for power-law galaxies 
discussed above. On the other hand, these objects are 
in the range of relatively low optical luminosity where few
radio-bright objects are found.

\section{Summary and conclusions}
\label{summary}
Aim of this series of three papers is to explore the connection
between the host galaxies and the AGN activity in nearby early-type
galaxies. The new developments in our understanding of the
relationship between the supermassive black-holes and galaxies
properties as well as in the structure of the central regions
of nearby galaxies warrant to re-explore this classical issue
in a new framework. In particular, early-type galaxies are the critical
class of objects. The two classes
defined on the basis of their brightness profiles 
(core and power-law galaxies) coexist in early-type galaxies  
and they host both radio-quiet and radio-loud AGN. 

A robust classification of the brightness profile can only be
obtained for nearby objects since the
most compact cores ($r \sim 10$ pc) are 
barely resolved even in the HST images already at a distance of 40
Mpc. We then focus on two nearby samples of early-type galaxies 
well studied at radio-wavelengths in the past and more specifically
on 116 galaxies (out a complete sample of 312 objects) that have been
detected above the 1 mJy threshold in VLA images at 5 GHz by
\citet{wrobel91a} and \citet{sadler89}. 
Imposing a radio-flux limit boosts the fraction of AGN with respect to a
purely optically selected sample and, at the same time, provides us
with a crucial information (their radio-luminosity) 
on the nuclear activity of these objects. Due to their proximity and
to the low radio-flux limit we will be able to explore the AGN
properties in objects with a radio-power as low as 
$L_r \sim 10^{19}$ W  Hz$^{-1}$ (or , equivalently, to $\nu L_r \sim 10^{36}$ erg s$^{-1}$).

In this first paper we analyze the archival HST images 
for this radio-selected sub-sample. They are available for 65 objects.
Having discarded 11 objects with a complex morphology,
we fit elliptical isophotes and derive the brightness profile
for the remaining galaxies. With only 3 exceptions these profiles are
well reproduced by a Nuker law fit. We are then able to separate
early-type 
galaxies on the  basis of the slope of their  nuclear  brightness
profiles ($\gamma$), into core ($\gamma \leq 0.3$) 
and power-law ($\gamma \geq 0.5$) galaxies;
we also found 3 galaxies with an intermediate slope
($ 0.3 < \gamma < 0.5$). We preferred
this classification with respect to  
the traditional morphological scheme (i.e. E and
S0 galaxies) since it is well known that it is often
difficult to  unambiguously distinguish between
these two classes of objects \citep{vandenbergh04}.

We recovered the already known difference in the luminosity between
core and power-law galaxies, since the most luminous galaxies 
are exclusively core galaxies. Below $M_K \sim -24.2$, however,
the two classes coexist and core galaxies are associated to objects
as faint as $M_K \sim -21.8$. A well defined correlation
between the radius and the surface brightness at the
profile break, $r_b$ and $\mu_b$, the ``core fundamental plane'', 
is present also in our sample and it is indistinguishable 
from that seen in the other samples, purely optically selected,
studied. Thus, our sample
does not reflect any significant difference in the host galaxies
with respect to the normal galaxies population.

Concerning the relationship
between host's magnitude and radio-power, 
we found that only core-galaxies are radio-emitters at level larger than
$L_r > 2.5 \times 10^{21}$ W Hz$^{-1}$, confirming the suggestions
by De Ruiter et al. (2005). But below this threshold
the two populations of early-type galaxies cannot be readily 
differentiated. Since no strong
radio-source ($L_r > 10^{22}$ W Hz$^{-1}$) is associated to a host
with $M_K > -24$ regardless of the profile class,
we are left with the ambiguity on the origin of this 
threshold in radio-luminosity
for power-law galaxies: it can be due both to a different 
nuclear structure of the two classes or to a
difference in the host galaxy's luminosity, since power-law
galaxies only scarcely 
populate the high end of the optical luminosity distributions, where
the brightest radio-sources are found.

The most significant difference concerning the optical
properties of our sample with respect to previous
studies is the higher fraction of nucleated galaxies, since
we detected an optical (or infrared)
nucleus in $\sim 65\%$ of the objects. 
This is not at all unexpected, since our sample is 
deliberately biased to favour the inclusion of active galaxies.
Addressing the multiwavelength properties of these nuclei will be the 
aim of the two forthcoming papers of the series.

\begin{acknowledgements}
This research has made use of the NASA/IPAC Extragalactic Database (NED)
(which is operated by the Jet Propulsion Laboratory, California Institute of
Technology, under contract with the National Aeronautics and Space
Administration), of the NASA/ IPAC Infrared Science Archive
(which is operated by the Jet Propulsion Laboratory, California
Institute of Technology, under contract with the National Aeronautics
and Space Administration) and of the LEDA database.
\end{acknowledgements}

\appendix

\section{Notes on the individual sources}
\label{notes}

{\bf UGC~5292}: the presence of a bright point source, combined with a
complex large scale profile prevents the fitting of this source.

\noindent
{\bf UGC~5663}: despite the presence of extended patches of dust, the
masking procedure allows us to derive its brightness profile 
which is well fitted by a Nuker law.

\noindent
{\bf UGC~6153}: dust patches cover a large fraction of the nuclear regions
of this galaxy and a bright point source is present. Nonetheless, 
the profile is relatively well behaved and suggests a classification 
as power-law. The residuals larger than
average for the other sources makes it tentative. 

\noindent
{\bf UGC~6297}: the brightness profile is well reproduced by a Nuker law,
despite the presence of dust filaments.

\noindent
{\bf UGC~7311}: the large scale edge-on dusty disk covers most of the galaxy for
$r<0\farcs7$. However, we derived a much larger break radius 
($r_b = 3.43\arcsec$) and concluded that the classification is robust.

\noindent
{\bf UGC~7614}: the presence of an inclined large scale disk induces 
relatively large fluctuations in the outer profile, but the relatively
unperturbed morphology at small radii allows us a robust classification
as a power-law galaxy.

\noindent
{\bf UGC~7575}: the situation for this object is similar to that of 
UGC~7614, with a large scale disk but little dust in the innermost
1\arcsec, leading to a clear  power-law classification.

\noindent
{\bf UGC~9723}: in this object an almost exactly edge-on dusty disk is present, but nonetheless 
the spheroidal component can be accurately traced, after masking
down to $\sim$ 2 \arcsec. The break radius derived from the Nuker fit is substantially
larger $\sim 3$\arcsec, suggesting a tentative core classification.

\noindent
{\bf NGC~1316}: despite the presence of extended patches of dust, the
brightness profile is remarkably smooth and well fitted by a Nuker law.

\noindent
{\bf NGC~1380}: after masking the off-center dust lane, the
brightness profile is smooth and well fitted by a power-law.
No flattening is seen
before the resolution limit.

\noindent
{\bf NGC~3258}: a circum-nuclear well defined dusty disk prevents the study of the
brightness profile for $r<0\farcs9$. The innermost points of the
profiles suggest the presence of a flat core. The best fit value is
found for  $r_b=1\farcs15$, i.e. apparently the profile breaks before
the edge of the disk. Forcing the value of $\gamma$ to 0.3 requires
a strong monotonic increase of dust obscuration toward the center; at $r<0\farcs1$
the surface brightness should be depressed by a factor 2.5, difficult
to reach considering the presence of starlight {\sl in front} of the disk. 
These considerations suggest a classification as a likely core galaxy.

\noindent
{\bf NGC~3268}: dust filaments cover the region between 0\farcs25 and
1\farcs2. Nonetheless the profile is characteristic of a core galaxy.
 
\noindent
{\bf NGC~3557}: the behaviour galaxy is very similar to NGC 3258, see above.
Here the break radius is at  $r_b=1\farcs60$, while the dusty disk has
a radius of $r=0\farcs95$. The brightness deficit at the center in this case
for  $\gamma=0.3$ would be of a factor 2.2, 
again suggesting a classification as a likely core galaxy.

\noindent
{\bf NGC~3706}: the central region of this galaxy has a peculiar morphology,
being extremely elongated. 
The brightness profile has a steep outer slope, which increases below
0\farcs4 (approximately the size of the elongated feature) then 
changing to a flat plateau. We conservatively classify this profile as
complex.

\noindent
{\bf NGC~5128}: the fit has been performed on a dust-corrected image
kindly provided by A. Marconi, obtained combining H and K band HST images
of this source. The central 1\arcsec\ is flagged due to
the presence of a strong point source.

\noindent
{\bf NGC~5419}: The central 0\farcs15 are flagged due to
the presence of a point source. A fainter second point source (flagged in the
fitting) is located close to the nucleus.

\noindent
{\bf NGC~6958}: this galaxy  shows a well defined power-law  decrease with a
constant  slope $\beta =  1.22$, after  excluding a  rather extended
region with a light excess  reaching 0.2 dex.  No flattening is seen
before the resolution limit. 

\noindent
{\bf IC~3370}: dust patches cover a large fraction of the nuclear regions
of this galaxy. The resulting profile is quite complex and no fit has
been performed.

\noindent
{\bf IC~4296}: the circum-nuclear dusty disk seen in the optical HST images
\citep{laine03} affects only marginally the NICMOS image
used by us. The central 0\farcs4 are flagged due to
the presence of a point source.

\end{document}